\documentclass{cjpsuppl}
\input epsf.tex
\input psfig.sty
\catcode`\@=11
\def\ps@fnal{\def\@oddhead{\textsf{FERMILAB--CONF--05/038--T \hfil \thepage}}
\def\@evenhead{\thepage \hfil \textsf{FERMILAB--CONF--05/038--T}}}
\relax
\begin{document}
\title{Phenomenology Beyond the Standard Model}
\authori{Joseph D. Lykken}    
\addressi{Theoretical Physics Dept.\\
Fermi National Accelerator Laboratory\\
P.O. Box 500, Batavia, IL 60510 USA}
\authorii{} \addressii{}
\authoriii{}   \addressiii{}
\authoriv{}    \addressiv{}
\authorv{}     \addressv{}
\authorvi{}    \addressvi{}
\headtitle{Phenomenology Beyond the Standard Model}
\headauthor{Joseph D. Lykken}
\lastevenhead{Joseph D. Lykken: Phenomenology Beyond the Standard Model}
\pacs{}
\keywords{}

\daterec{}


\maketitle
\begin{abstract}
An elementary review of models and phenomenology for physics beyond
the Standard Model (excluding supersymmetry). The emphasis
is on LHC physics. Based upon a talk given
at the \textit{Physics at LHC} conference, Vienna, 13-17 July 2004.
\end{abstract}

\section{Causarum Investigatio}

Written on the ceiling fresco of the beautiful Festsaal of the
\"Osterreichischen Akademie der Wissenschaften, are two words:
Causarum Investigatio. Just as these words must have inspired
Boltzmann and Schr\"odinger, today we are inspired to investigate
the causes and more fundamental structures underlying the Standard Model of
particle physics.

The Standard Model is in remarkably successful agreement with particle
physics data at all energies below the scale of electroweak symmetry
breaking. Indirect probes of physics up to multi-TeV scales, using
rare decays, or using the sensitivity of electroweak precision data to
virtual processes, also show no significant deviations. Thus, in a
strict experimental sense, we have no guidance for moving beyond the
Standard Model, other than the clear anomalies of neutrinos oscillations,
dark matter, and dark energy.

\begin{figure}
\centerline{\epsfxsize 3.75 truein \epsfbox {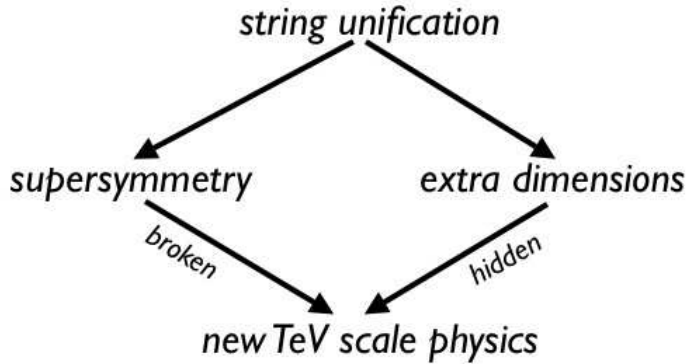}}
\caption{\label{fig:bigpic} The big picture of physics
beyond the Standard Model.}
\end{figure}

However, the success of the Standard Model indicates that fundamental
physics is closely tied to the basic principles of quantum mechanics
and to symmetry principles, of which the most notable are relativity
and local gauge invariance. These provide a very constrained set of rules
for extensions of the Standard Model, and a great deal of theoretical
investigation in the past twenty five years has been devoted to mapping
out the possible scenarios consistent with these rules.

Indeed, our theoretical understanding of these structures is now so mature
that it is not overly pretentious to sketch the ``big picture'' of physics
beyond the Standard Model. This sketch is illustrated in 
Figure \ref{fig:bigpic},
where the vertical direction represents energy scale. The two scales
shown are: the ``TeV'' scale characterized by the new physics responsible
for electroweak symmetry breaking, and the ``string unification'' scale,
defined as the scale where one begins to have a unified description of
quantum gravity with the gauge interactions of the Standard Model.

We know the TeV scale (within an order of magnitude), but we do not know
what is the new physics responsible for electroweak symmetry breaking.
Many very distinct and theoretically viable mechanisms have been proposed.
Identifying and understanding this physics is the primary goal of the
LHC project.

We do not know the scale of unification (even to within an
order of magnitude) nor do we know to what extent
it involves gauge coupling unification, grand unification, flavor
unification, or superstrings. But our best guess is that some combination
of all of these elements is involved. Determining this scale and
uncovering the new physics in operation there is one of the ultimate
goals of particle physics.

The figure also demonstrates another equally important and more
immediate goal. If unification occurs in any form, then there must be
highly sophisticated dynamical mechanisms which convert the simple
unified theory at ultra high energies into the messy junk which we
observe in experiments today. These are shown in the figure as the
mechanisms which respectively break supersymmetry and hide extra
dimensions. We do not understand either of these mechanisms, though
again theorists have proposed many distinct possibilities.
Determining which mechanisms Nature has chosen is a process we
expect to begin at the LHC, and continue with future colliders.

Since supersymmetry (SUSY) is well-covered in other sessions at the
conference \cite{Chung:2003fi},
I will concentrate here on extra dimensions and related
ideas. However I hasten to add that one should never regard
supersymmetry and extra dimensions as mutually exclusive. On the
contrary, it is extremely difficult to imagine that any of the
current ambitious schemes with extra dimensions can stably devolve
from the unification scale without the help of SUSY. It is
only somewhat easier to imagine that the minimal picture of
supersymmetric unification actually works, without at least some
assistance near the high scale from an orthogonal organizing principle
like extra dimensions. And we should not forget that
string theory needs them both.

It is also true that there are many topics in Beyond the Standard Model
(BSM) phenomenology which are neither SUSY nor extra dimensions.
However this division has grown fuzzy of late. For example,
almost all of the current theoretical research on models with new
strong dynamics are exploiting the AdS/CFT correspondence to map
these gauge theories into models of extra dimensions with branes.
Thus while our review contains no mention of technicolor, I will
discuss recent work on ``Higgsless'' models in extra 
dimensions \cite{Csaki:2003dt}.
Similarly much of the increase in our understanding of strongly
coupled gauge theories during the past decade has come from the
study of SUSY gauge theories, and this has begun to be reflected
in BSM model-building. Thus for example the recent ``fat Higgs''
models \cite{Harnik:2003rs}
modify the conventional SUSY desert scenario via an
elegant matching of two gauge theories while preserving gauge
coupling unification. This has the attractive feature of solving
the post-LEP ``little hierarchy'' problem, and at the very least
is a counterexample to arguments which purport to demonstrate the
unique attractiveness of the minimal SUSY scenario.
 
Similarly, what used to be called ``excited fermions'' in BSM
search talks now appear as ur-Kaluza-Klein modes in generic models of
deconstruction. Leptoquarks, another old warhorse of BSM searches,
will be discussed later in this review. I will show how leptoquarks
can most profitably be regarded as products of either SUSY or
extra dimensions.

Having thus dramatically shortened the usual laundry list for BSM reviews,
I end the introduction with an additional disclosure. This review will
be completely collider-centric. This is a clear deficiency,
especially in an era where the interconnectedness of high energy
physics and astrophysics plays an absolutely essential role. As
already mentioned, the only clear deviations from the Standard Model
to date are neutrino oscillations, dark matter, and dark energy. None
of these were discovered at colliders. In the future, I expect that
studies of rare processes, B physics, the 
neutrino sector, cosmology, particle astrophysics, and particle astronomy
will all provide important clues to BSM physics. However
I also expect collider experiments to be the major contributors to
reaching the ambitious goals which I have outlined above.

\section{A bestiary of extra dimensions models}

BSM review talks 15 years ago usually made no mention at all of
extra dimensions. Since all of the reasons for taking extra dimensions
seriously existed 15 years ago, this was a purely sociological
absence, i.e., extra dimensions were not socially acceptable. This is perhaps
a residual effect of the curse of Gunnar Nordstr\"om, the Finnish
physicist who invented Kaluza-Klein theory in 1914, only to have
his brilliant idea completely ignored by
Einstein. Kaluza-Klein theory was, in turn, ignored by almost all physicists
for a half century, finally being resuscitated by purveyors of supergravity
and superstrings and the 1970s and 80s.

Briefly, there are three strong motivations for attempting to incorporate
extra dimensions in BSM physics. The first motivation is the Standard
Model (SM) itself, which has too many elementary particles (57 not including
the Higgs) for a theory of fundamental constituents. Especially when one
factors in the enormously complex flavor structure of the SM,
it is clear that new dynamics is at work here, involving new degrees of
freedom intimately connected to the SM degrees of freedom. New gauge
interactions with resulting composites may be part of the answer, but
probably not all of it. This leaves only two other known directions: extra
dimensions, which when dynamically compactified or otherwise hidden create
complex low energy patterns, and extended objects, which leads to string
theory. String theory is itself the second prime motivation for extra
dimensions, since quantized strings do not give reasonable physics unless
they are embedded into a ten-dimensional spacetime. The third motivation
is quantum gravity, broadly defined, which tells us that space and time
are to be regarded as themselves fully ``dynamical'' objects. While it is
not clear exactly what this means, it certainly implies that the
number of accessible spatial dimensions at a given energy scale should
be regarded as a dynamical physical observable, not given {\it a priori}.

Depending upon the physical mechanism invoked to hide extra dimensions
from current observation, there is a great range of possible
energy or length scales at which they may begin to appear.
The most conservative guess is that their inverse radii are within about
an order of magnitude of the unification scale; even in this case
extra dimensions can have very significant effects upon physical
observables at the TeV scale. In many models, the extra dimensions
themselves appear around the TeV scale, and are linked to TeV scale
physics such as electroweak symmetry breaking (EWSB) or supersymmetry
breaking. In models which employ very efficient brane methods for
hiding extra dimensions, the extra dimensions can even be of macroscopic
size without contradicting any current observations or experiments.
Indeed an extra dimension of infinite extent is not necessarily excluded.

A complete bestiary of extra dimensions models is beyond the scope of
this review. We can make a partial survey based upon a simple organizing
principle: what is the physical mechanism which hides the extra dimensions?
Possible answers include:
\begin{itemize}
\item The extra dimensions are compact and small. Examples include a circle, a
sphere, a torus, a Calabi-Yau manifold, and various kinds of orbifolds.
\item Some or all of the SM particles are confined to a brane or an
intersection of branes, and thus cannot probe the full 
extra dimensional ``bulk'' space.
\item The extra dimensions are fundamentally different. For example,
if the extra dimensions are fermionic, we are back to supersymmetry.
The extra dimensions might also be discrete or otherwise
``deconstructed'', so that they only approximately resemble spatial
degrees of freedom in a certain energy regime.
\item Any combination of the above.
\end{itemize}
In addition, it is important to specify whether the extra dimensions
are curved or flat, and whether the various bulk fields have nontrivial
vacuum configurations in the extra dimensions.

A glance at SPIRES reveals that there are in excess of 3,000 papers
discussing extra dimensions models which fit into the above
categorization. I can summarize the current status of these extra
dimensions models in two bullets:
\begin{itemize}
\item There are too many models.
\item None of them are any good.
\end{itemize}
The first bullet is obvious to any experimenter interested in searching
for signals of extra dimensions at the LHC. What models should we
simulate? What are the key phenomenological features and key
discriminators between different models? These are basic questions
which need to be answered before LHC turn-on.

The second bullet brings up some points which further complicate
our task. Most of the work on extra dimensions is at the level which
Savas Dimopoulos aptly calls ``scenarios'' rather than models. A
scenario is defined (by me) as a set of physical assumptions which,
with considerably more work, could evolve into a respectable class of
models. Many of these scenarios have remained scenarios because they
suffer from deep theoretical problems or ``gaps'', and some of them
have fairly nasty (but generic) phenomenological problems. This is
not a reason to denigrate these scenarios -- after all the same could
be said of supersymmetry models or of string theory! But it does
explain why we are still some way from having a decent set of 
serious and ``theoretically stable'' benchmark
models for simulation. 

That said, I will review the status of three out of the four
classes of extra dimensions scenarios which are are most relevant
for LHC searches. The fourth class, direct compactifications of
string theory, is potentially the most interesting, but as of 2004
was not quite ready to make contact with 
LHC physics \cite{Kane:2004hm}-\cite{Kobayashi:2004ya}.

\section{UED}

Universal Extra Dimensions is the name given by Appelquist, Cheng, and
Dobrescu \cite{UED}
to a class of models which closely resemble the original
Nordstr\"om-Kaluza-Klein scenario, but with some crucial improvements.
All particles live in the full bulk, which is compactified to some
kind of orbifold. The simplest case is a single extra dimension
with coordinate $y$,
compactified to a circle, which in turn is orbifolded to a line
interval of length $L$ by identifying points under $y\to -y$. The orbifolding
is necessary because otherwise the Kaluza-Klein zero modes
of fermions (i.e. the light 4d fermions) are vectorlike.
The orbifold projections in the above example remove half of the
chiralities for the fermion zero modes,
allowing an effective 4d theory which matches
the chiral Standard Model.

These same orbifold projections have other good effects.
For example, consider a 5-dimensional bulk gauge field 
$A_M = (A_{\mu},A_y)$, where $\mu$ is a 4d vector label and $y$ labels
the fifth dimension. Since $A_y$ appears in a 5d covariant derivative
with $d/dy$, we keep the odd Kaluza-Klein (KK) modes of
$A_y$ after orbifolding, while keeping the even KK modes of the $A_{\mu}$.
This means that $A_{\mu}$ has a massless zero mode, but $A_y$ does
not. Thus we manage to avoid having a massless adjoint scalar 
accompany every massless gauge boson in the effective
4d theory.

In the original Kaluza-Klein model Kaluza-Klein mode number
is conserved, as this is just conservation of (discrete) momentum
in the extra dimension. However in
our simple UED example there are two orbifold fixed points:
one is $y=0$ and the other is $y=L$. Thus orbifolding breaks
the translational invariance of the circle, by distinguishing two
special points. This may have no effect at tree level, but radiative
corrections will generically introduce interactions which violate
conservation of KK mode number.
However a single remnant translation,
$y\to y+L$, is still a symmetry, since it just interchanges the
two fixed points. Since translation invariance is broken to
a $\mathbf{Z_2}$ remnant, momentum conservation in the fifth dimension
is replaced by a conserved parity, called ``KK parity''.
Zero modes are even under this parity but the lightest massive
modes are odd.

This is enough to guarantee that the lightest massive KK mode
in a UED model is stable. The situation is quite analogous
to R parity in SUSY models. As with SUSY, this implies that
in UED models the first massive KK modes must be produced in pairs.
More generally, production of any single massive KK mode is quite
suppressed, unless one introduces large tree level couplings which
violate KK momentum conservation.
This in turn suppresses the virtual corrections of these KK modes to
Standard Model processes, allowing the UED scale $1/L$ to
be as low as 300 GeV before we get into conflict with
precision data. It also means that the lightest massive KK
mode, the ``LKP'', is a good cold dark matter 
candidate \cite{Cheng:2002ej,Servant:2002aq}.

\begin{figure}
\centerline{\epsfxsize 3.75 truein \epsfbox {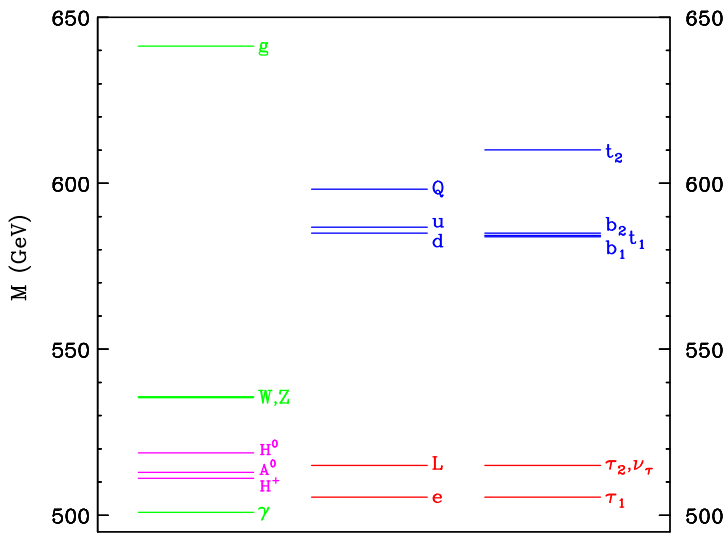}}
\caption{\label{fig:KKlevels} Spectrum of the first massive
KK modes of the Standard Model particles, in a simple UED
model. Taken from reference \cite{Cheng:2002ab}.
\hbox to170pt{}}
\end{figure}

UED models, in addition to being rather simple, have fairly
universal predictions for colliders. If the LKP is a major constituent
of dark matter, then it is in a mass-coupling range such that it
will be produced at the LHC. Figure 2 shows a typical spectrum for
the first massive KK modes in a UED model \cite{Cheng:2002ab},
after taking into account
the mass splittings from radiative corrections. As in SUSY models
the partner of the SM gauge boson $B$ tends to have the smallest
radiatively corrected mass. The other first massive KK modes will
decay promptly to this LKP. Thus typical UED events at the LHC
give a variety of jet and lepton signatures combined with large
missing transverse energy (MET).

If you only produce the first massive KK modes, UED models look
very much like a subset of SUSY models, in terms of their collider
signatures. Even if you detect a few of the second level KK
modes, it is not obvious that this will 
dramatically disambiguate the signatures 
from an extended SUSY model; this should be studied.
The crucial discriminators, of course, are the spins of
the heavy partner particles. Distinguishing these spins is a very
significant experimental challenge \cite{Barr:2004ze}.

If we are lucky and the UED scale is close to the current bound,
i.e. $1/L \sim 300$ GeV, then it will be possible to see UED
effects as mimimal flavor violating loops in 
heavy flavor physics \cite{Buras:2003mk}.

\section{ADD}

ADD is the name I use to refer to the class of models which incorporate
the large extra dimensions scenario of Arkani-Hamed, Dvali, and
Dimopoulos \cite{Arkani}.
These were the first extra dimensions models in
which the compactified dimensions can be of macroscopic
size, consistent with all current experiments and observations.
For this reason they are sometimes known as ``large extra dimensions''
models.

In the most basic version, $n$ extra spatial dimensions are
compactified on a torus with common circumference $R$, and
a brane is introduced which extends only in the three infinite
spatial directions. Strictly speaking, the brane should have
a very small tension (energy per unit volume) in order that
it does not significantly warp the extra dimensional space.
It is assumed that all of the SM fields extend only in the brane.
This can be considered as a toy version of what happens
in string theory, where chiral gauge theories similar to the
SM are confined to reasonably simple brane configurations in
reasonably simple string compactifications \cite{Lykken:1998ec}.

An immediate consequence of these assumptions is that the
effective 4d Planck scale is related to the underlying
fundamental Planck scale of the $4$$+$$n$-dimensional theory
and to the volume of the compactified space. This relation follows
immediately from Gauss' Law, or by dimensional truncation:
\begin{eqnarray}
M_{\rm Planck}^2 = M_*^{2+n}\,R^n \quad ,
\label{eqn:gaussrel}
\end{eqnarray}
where $M_{\rm Planck}^2$ is defined by Newton's
constant: $M_{\rm Planck} = 1/\sqrt{G_N} = 1.2\times 10^{19}$ GeV. 
$M_*^{2+n}$ is defined as the gravitational coupling which
appears in the $4$$+$$n$-dimensional version of the Einstein-Hilbert
action. It is the quantum gravity scale of the higher dimensional theory.

If $M_{\rm Planck}$, $M_*$ and $1/R$ are all of the same order,
as is usually assumed in string theory, this relation is not very
interesting. But there is nothing which prevents us from assuming
that $M_*$ is equal to some completely different scale. Most
attractive is to take $M_* \sim 1$ TeV, and attempt \cite{Antoniadis:1990ew}
to replace
the hierarchy problem of the SM by a large compactification radius,
i.e. to swap an ultraviolet problem for an infrared one!
Note that, if we want to mantain contact with string theory,
ADD-like models must arise from string ground states in which the
string scale (and thus the ultraviolet cutoff for gravity)
is also in the TeV range. This is difficult but 
not unthinkable \cite{Lykken:1996fj}.

The ADD scenario raises the exciting possibility of observing
quantum gravity at the LHC. In such models only the graviton,
and possibly some non-SM exotics like the right-handed neutrino,
probe the full bulk space. There is a Kaluza-Klein tower of
graviton modes, where the massless mode is the standard 4d
graviton, and the other KK modes are massive spin 2 particles
which also couple to SM matter with gravitational strength.

Whereas bremstrahhlung of ordinary gravitons is a completely
negligible effect at colliders, the total cross section to
produce {\it some} massive KK graviton is volume enhanced,
and thus effectively suppressed only by powers of $M_*$,
not $M_{\rm Planck}$. From Eq. (\ref{eqn:gaussrel}) one obtains:
\begin{eqnarray}
\sigma \sim {1\over M_{\rm Planck}^2}(ER)^n
\sim {1\over M_*^2}(EM_*)^n \quad ,
\end{eqnarray}
where $E$ is the characteristic energy of the subprocess.

For graviton phenomenology it is useful to replace
the ADD parameter $M_*$ by other
rescaled parameters. The two most useful choices are taken from the
work of Giudice, Rattazzi and Wells (GRZ) \cite{Giudice:1998ck}, and
Han, Lykken and Zhang (HLZ) \cite{Han:1998sg}:
\begin{eqnarray}
M_*^{n+2} &=& {S_{n-1}\over (2\pi )^n} M_s^{n+2} \; ,\\
M_*^{n+2} &=& {8\pi\over (2\pi )^n} M_D^{n+2} \; ,
\end{eqnarray}
where $M_s$ is the HLZ scale, $M_D$ is the GRW scale, and $S_{n-1}$ is
the surface area of a unit $n$-sphere:
\begin{eqnarray}
S_{n-1} = {2\pi^{n/2}\over\Gamma ({n/2})} \; .
\end{eqnarray}
Both notations are equivalent.
To obtain a complete dictionary between ADD, GRZ and HLZ,
one also needs to relate the ADD parameter $R$ to those used by
the other authors:
$R = R_{\rm HLZ} = 2\pi R_{GRW}$, 
and take note of the
different notations for Newton's constant:
\begin{eqnarray}
\kappa^2 = 16\pi G_N \; (\rm HLZ);\quad 
\bar{M}_P^2 = {1\over 8\pi G_N} \; (\rm GRW)\; .
\end{eqnarray}

A Kaluza-Klein (KK) graviton mode has a mass specified by an
$n$-vector of integers $\vec{k}$:
\begin{eqnarray}
m^2(\vec{k}) = {\vec{k}^2\over R_{\rm GRW}^2} \; .
\end{eqnarray}
Let $r=\vert\vec{k}\vert$. Then for large $r$ (as is always the
relevant case for ADD phenomenology) the number of KK
graviton states of a given polarization with $r\le r_{\rm max}$ is
given by the integral
\begin{eqnarray}
S_{n-1}\int_0^{r_{\rm max}}dr\, r^{n-1} 
&=& {1\over n}S_{n-1}\; r_{\rm max}^n \; \nonumber \\
&=& \int_0^{m_{\rm max}} \rho (m) \,dm \; ,
\end{eqnarray}
where the KK density of states is
\begin{eqnarray}
\rho (m) = {m^{n-1}\over G_N M_s^{n+2}} \; .
\end{eqnarray}
We see that $M_s$ is the natural scaling parameter for KK graviton
production. The density of states formulation can be applied to a much
more general class of models than ADD, and can also include
graviton wavefunction factors when the extra dimensions are not flat.

Consider now on-shell production of a KK graviton from a $pp$ or
$p\bar{p}$ collision. To leading order this is a $2\rightarrow 2$
process with two massless partons in the initial state, plus a massive
KK graviton and a massless parton in the final state. Let
$p_1$, $p_2$ denote the 4-momenta of the
initial state partons, $p_3$ the 4-momentum of the
graviton, and $p_4$ the 4-momentum of the outgoing parton.
The total cross
section for any particular variety of partonic subprocess has the form
\begin{eqnarray}
\sigma (1 + 2 \rightarrow {\rm KK} + 4) =
\int dx_1dx_2\, f_1(x_1,\hat{s})f_2(x_2,\hat{s})\int d\hat{t}
\int_0^{\sqrt{\hat{s}}} \hspace{-7pt}dm \, \rho (m)\, 
{d\sigma_m\over d\hat{t}}(\hat{s},\hat{t})\, ,
\end{eqnarray}
where $f_1(x_1,\hat{s})$, $f_2(x_2,\hat{s})$ 
are the parton distribution functions
(pdfs) for the intitial state partons, 
$\hat{s} = x_1x_2s = (p_1 + p_2)^2$ is the
square of the total center of mass (cm) energy of the subprocess, and
$\hat{t} = (p_1 - p_3)^2$ is the usual Mandelstam invariant. The
formulae for $d\sigma_m/d\hat{t}$, the differential subprocess cross
sections for KK gravitons of mass $m$, are given in equations 64-66 of
GRW.

\section{RS}

Randall-Sundrum refers to a class of scenarios, also known as
warped extra dimensions models, originated by Lisa Randall
and Raman Sundrum \cite{Randall:1999ee,RSII}.
In these scenarios there is one extra spatial
dimension, and the five-dimensional geometry is ``warped'' by the
presence of one or more branes. The branes extend infinitely in the usual
three spatial dimensions, but are sufficiently thin in the warped direction
that their profiles are well-approximated by delta functions in the energy
regime of interest. If we ignore fluctuations of the branes, we can always
choose a ``Gaussian Normal'' coordinate system, such that the fifth dimension
is labelled $y$ and the usual 4d spacetime by $x^{\mu}$. The action for
such a theory contains, at a minimum, a 5d bulk gravity piece and
4d brane pieces. The bulk piece has the 5d Einstein-Hilbert action with
gravitational coupling $M^3$, and a
5d cosmological constant $\Lambda$.
The brane pieces are proportional to the brane tensions $V_i$,
which may be positive or negative. These act as sources for 5d gravity,
contributing to the 5d stress-energy terms proportional to
\begin{eqnarray}
\sum_iV_i\delta(y-y_i)
\end{eqnarray}
where the $y_i$ are the positions of the branes. 
Combined with a negative $\Lambda$, this results in a curved
geometry, with a 5d metric of the form:
\begin{eqnarray}
g_{\mu\nu}(x^{\rho},y) = a^2(y) \,{\tilde{g}}_{\mu\nu}(x^{\rho})\; ,
\nonumber\\
g_{\mu y} = 0\; ,\quad g_{yy} = 1 \; ,
\end{eqnarray}
where $a(y)$ is called the warp factor, $\tilde{g}$ is a 4d metric, and
I have made a useful choice of coordinates. Warping refers to the fact that
a 4d distance $d_0$ measured at $y=y_0$ is related to an analogous 4d distance
$d_1$ measured at $y=y_1$ by $a(y_0)d_0 = a(y_1)d_1$. Thus in Randall-Sundrum
scenarios 4d length, time, energy and mass scales vary with $y$.

So far we are being completely general. However almost all collider physics
phenomenology done with warped extra dimensions so far is based upon
one very specific model, the original simple scenario called RSI.
In this model the extra dimension is compactified to a circle of
circumference $2L$, and then further orbifolded by identifying
points related by $y\to -y$. Thus the fifth dimension consists of two
periodically identified mirror copies of a curved 5d space
extending from $y=0$ to $y=L$. It is assumed that there is a brane
at $y=0$, with positive tension $V_0$; it is known as the Planck brane.
There is another brane at $y=L$, with negative tension $V_L$, known
as the TeV brane. 

Randall and Sundrum showed that, for a specially tuned
choice of input parameters $V_0 = -V_L = -M^2\Lambda$, the 5d Einstein
equations have a simple warped solution on $0 < y < L$ with metric:
\begin{eqnarray}
g_{\mu\nu}(x^{\rho},y) = {\rm e}^{-2ky} \,\eta_{\mu\nu}\; ,
\nonumber\\
g_{\mu y} = 0\; ,\quad g_{yy} = 1 \; ,
\end{eqnarray}
where $\eta_{\mu\nu}$ is the 4d flat Minkowski metric, and
$k = \sqrt{-\Lambda}$. Away from the branes, the
5d curvature is constant and negative; it is thus equivalent locally to
$AdS_5$, with the Anti-de Sitter radius of curvature given by
$1/k$. At the locations of the branes the curvature is discontinuous, due
to the fact that the branes are delta function sources for curvature.  

We see that the RSI model is completely described by three parameters:
$k$, $M$, and $L$. Since we are not resolving the brane profiles, and
since we do not want to worry about how to embed this scenario into
string theory or some other description of quantum gravity, we had
better restrict ourselves to a low energy effective description. This implies
taking $k$, $1/L \ll M$. In fact in RSI it is assumed that $k$ is
merely parametrically small compared to the 5d Planck scale $M$, i.e.
something like $k \sim M/10$. The effective 4d Planck scale, which
is the same thing as the coupling of the graviton zero mode, is
given by dimensional truncation: 
\begin{eqnarray}
M_{\rm Planck} = {M^3\over 2k}\left( 1-{\rm e}^{-2kL} \right) \; .
\end{eqnarray}
Thus, within an order of magnitude, $M \sim k \sim M_{\rm Planck}$.
In RSI we fix the distance $L$ by requiring that 
$a(L)M_{\rm Planck} \simeq 1$ TeV, thus $kL \sim 30$. This is
{\it not} a large extra dimension: its inverse size is comparable
to the grand unification scale.

The RSI model makes one further simple but drastic assumption: that
our entire 4d universe is confined to the TeV brane. From a particle
physics viewpoint (and completely ignoring cosmology) this amounts to
the statement that the Standard Model fields live on the TeV brane.
Thus, as in ADD models, the phenomenology of RSI is concerned with the
effects of the massive KK modes of the graviton. These modes are
quite remarkable since, as measured on the TeV brane, their mass
splittings are on the order of a TeV, and their couplings to SM
fields are only TeV suppressed. In RSI, the Standard Model is replaced
at the TeV scale by a new effective theory in which gravity is still
very weak, but there are exotic heavy spin-two particles.

\begin{figure}
\centerline{\psfig{figure=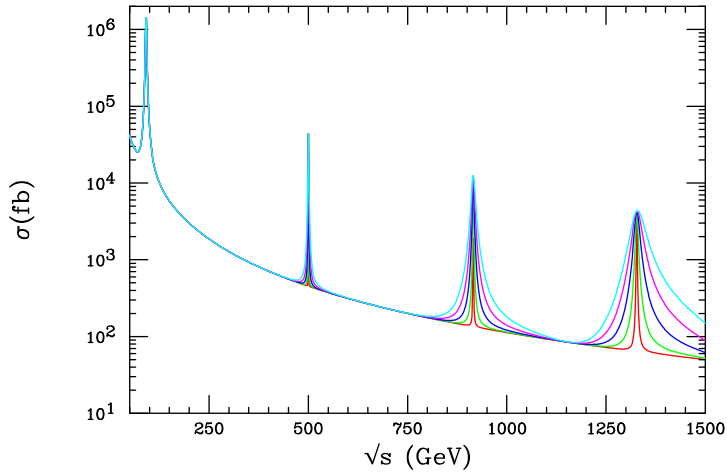,width=3.75 truein,angle=90}}
\caption{\label{fig:rsres} The cross section for
$e^+e^-\to \mu^+\mu^-$ including the exchange of KK gravitons
in the RSI model. The narrowest resonances correspond
to $k/M = 0.05$, the widest to $k/M = 0.14$.
Taken from reference \cite{Hewett-Smaria}.\hbox to200pt{}}
\end{figure}

At the LHC the KK gravitons of RSI would be seen as difermion or
diboson resonances, since (unlike the KK gravitons of ADD) the coupling
of each KK mode is only TeV suppressed \cite{Davoudiasl:1999jd}.
The width of these resonances
is controlled by the ratio $k/M$; the resonances become more
narrow as $k/M$ is reduced, as shown in Figure \ref{fig:rsres}.

\section{A behaviorist approach to extra dimensions}

The three wildly popular scenarios just reviewed are probably all too
simplistic to stand as viable candidates for top-down theories of extra
dimensions. But as already mentioned, our only viable top-down theory
of extra dimensions, string theory, is not yet understood well enough to
guide experiments at the LHC.

For collider phenomenology, we don't care so much where the extra dimensions
models came from; what we care about is what they do. This suggests a more
bottom-up approach, where we don't worry (much) about whether our models can
be fleshed out into globally respectable theories. Instead, we focus on
particular theoretical conundrums which can be solved by invoking extra
dimensions. Then we ask what are the distinctive phenomenological features
of each solution.

Here is a partial list of ``solutions'' to theoretical problems 
which have been suggested using extra dimensions
(often in more than one way):
\begin{itemize}
\item explain or assist electroweak symmetry breaking
\item explain the little hierarchy problem
\item lower the Planck scale 
\item break supersymmetry
\item explain flavor properties of the SM
\item improve grand unification
\item explain neutrino physics
\item explain dark matter
\item explain dark energy
\end{itemize}
I will briefly review three such phenomenologically oriented efforts:
little Higgs, Higgsless models, and
asymmetrical extra dimensions.

\section{Deconstructing little Higgs}

The SM is an effective field theory with a cutoff scale $\Lambda$. Since the
Higgs has quadratically divergent radiative corrections at one loop, we
expect that $\Lambda \sim (4\pi/g)*174$ GeV $\sim 1$ TeV. It is therefore
surprising that precision collider tests, with multi-TeV sensitivity,
see no evidence for any of the
dimension 5 and 6 operators which can be constructed purely from Standard
Model fields, obeying all SM symmetries. This is known as the little
hierarchy problem, since taken at face value it suggests that the cutoff
$\Lambda$ may be closer to 10 TeV than it is to 1 TeV.

Little Higgs models \cite{Arkani-Hamed:2002qy}-\cite{simplest}
address this problem by making the Higgs a
pseudo-Goldstone boson of new global symmetries which are both
explicitly broken (by SM gauge interactions) and spontaneously broken.
Roughly speaking, this buys you another factor of $4\pi$ in the relation
given above, allowing the cutoff $\Lambda$ to be naturally around 10 TeV.

Little Higgs model builders will claim that their models have nothing to
do with extra dimensions, but as far as I know extra dimensions are the
best motivation for this scenario.

Consider a 5d $SU(N)$ gauge theory. Force the extra dimension to be discrete,
i.e. a finite periodic lattice with $m$ sites and lattice spacing $1/f$.
The 5d Yang-Mills lagrangian then truncates to
\begin{eqnarray}
{1\over 2g^2}\sum_{i=1}^m{\rm tr}\,F_i^2
+ f^2\sum_{i=1}^m{\rm tr}\,\left[(D_{\mu}U_i)^{\dagger}D^{\mu}U_i\right]\;.
\end{eqnarray} 
For finite $m$ this is really a 4d theory, with $m$ different 
sets of $SU(N)$ gauge bosons,
together with ``link'' scalars $U_i$ 
which are bifundamentals, i.e. they each carry charges under two different
``adjoining'' $SU(N)$'s. The $U_i$ are just the latticized versions of
the $A_y$, the extra-dimensional components of the Yang-Mills field, which
from the 4d point of view are scalars in the adjoint representation.
If the $|U_i|$ get equal vevs, the gauge symmetry
will be broken down to a single diagonal $SU(N)$. The $U_i$ scalars provide
Goldstone bosons which make all but one combination of the $m$ sets of gauge
bosons massive. The spectrum of this 4d theory thus approximates the
KK spectrum of a true 5d Yang-Mills theory. Such 4d theories are called
deconstructed. They are obviously a much larger class of models than
the usual extra dimensional constructions.

In this example one scalar mode, contained in the product
$U_1U_2\ldots U_m$, is not eaten. It remains as a naturally light
pseudo-Goldstone mode. This is a simple example of a little Higgs.
The price we pay is that there are additional exotics with masses of
order the inverse lattice spacing $f$. These include (at least)
heavier gauge boson copies of the $W$, $Z$, or $B$, along with heavy exotic
scalars which are triplets or singlets under $SU(2)_L$. The extra
scalars occur because we are trying to extract the doublet Higgs of the
SM from the adjoint representation of something. Further complications
ensue incorporating the SM fermions, and one is forced at a minimum to
add a heavy exotic charge 2/3 
weak singlet quark $T$, which can be thought of as the
vectorlike partner of the right-handed top quark. 

Little Higgs models can be constructed to implement a conserved
quantum number called T-parity \cite{Cheng:2004yc,Low:2004xc}.
T-parity is the analog of KK parity
in UED models, and R-parity in SUSY models. Conserved T-parity implies
that the heavy exotics must be produced in pairs. This eliminates
tree level contributions from these exotics to precision electroweak
observables, allowing the fundamental scale $f$ to be as low as
$\sim 500$ GeV without contradicting experiment. The lightest exotic
with odd T-parity is stable and likely to be the $B^{\prime}$;
this is a viable cold dark matter
candidate \cite{Hubisz:2004ft,Cheng:2004yc}.
With conserved T-parity, Little Higgs models predict missing
energy signatures at the LHC, similar to both SUSY and 
UED \cite{Hubisz:2004ft}.

\begin{figure}
\centerline{\epsfxsize 3.75 truein \epsfbox {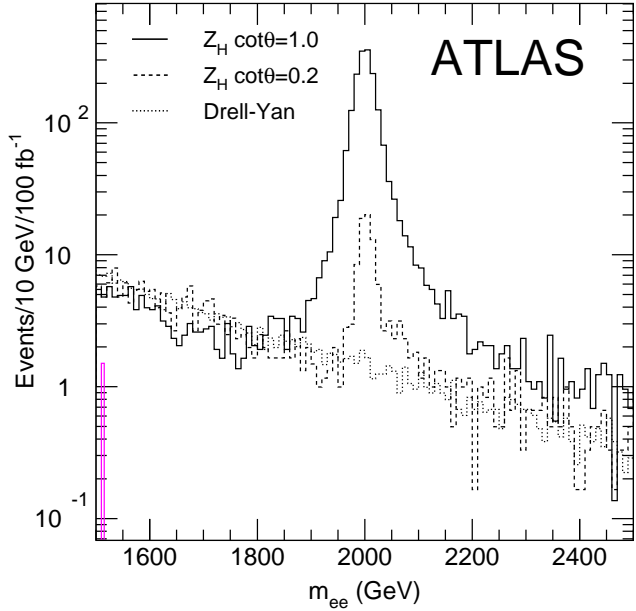}}
\caption{\label{fig:zprime} The peak in the $e^+e^-$ invariant mass
distribution from production of the  $Z^{\prime}$ of Little Higgs
with mass 2 TeV. The larger/smaller peaks correspond to
smaller/larger values of a mixing angle. Taken from
reference \cite{Azuelos:2004dm}.\hbox to200pt{}}
\end{figure}

Little Higgs models without a conserved T-parity are strongly constrained
by electroweak precision data \cite{Csaki:2002qg}-\cite{Casalbuoni:2003ft}. 
The lower bound on the scale $f$ is
in the range from 1 to 4 TeV. 
As seen in Figure \ref{fig:zprime} the $Z^{\prime}$ is
probably observable at the LHC as a peak in Drell-Yan 
production \cite{Azuelos:2004dm,Han:2003wu}.
More importantly, the decay $Z^{\prime} \to Zh$, which distinguishes
Little Higgs from other models with a $Z^{\prime}$, is probably
observable in $Z +$ b-jet channels over the same kinematic 
range \cite{Azuelos:2004dm,Burdman:2002ns}.
Single production
of the $T$ fermion is observable for masses up to and perhaps exceeding
1 TeV \cite{Azuelos:2004dm,Han:2003wu,Allanach:2004ub}.  

\section{Higgless models}

Let's modify the UED scenario by requiring the Higgs scalar field
to be localized in the fifth dimension to the orbifold fixed point
$y=L$. The electroweak gauge bosons inhabit the entire bulk, and
for simplicity we will denote them by a single 5d gauge coupling
$g_5$ and a
single gauge index: $A^a_M(x,y)$.
The action for the Higgs is then:
\begin{eqnarray}
\int d^4xdy\,\delta(y-L)\, \left(
{1\over 2}D_{\mu}\Phi_i D^{\mu}\Phi_i -V(\Phi )\right) \; ,
\end{eqnarray}
where $D_{\mu}\Phi_i = \partial_{\mu}\Phi_i
+g_5A^a_{\mu}T_{ij}^a\Phi_j$. In this covariant derivative
the 5d gauge coupling has mass dimension $-1/2$:
$g_5^2 = g_4^2L^2$.
This is compensated
by the fact that in 5d the bulk gauge field $A^a_{\mu}$ has
canonical mass dimension $3/2$.

If we derive the equations of motion (EOM) for this theory,
we will get a funny extra delta function piece in the
equation for the gauge field. Away from $y=L$ we
get just the EOM of the bulk gauge theory. Integrating  
the EOM in $y$ then picks up the contribution from the
delta function:
\begin{eqnarray}
\partial_y A^a_{\mu}(x,L) = g_5^2v^2A^a_{\mu}(x,L) \; ,
\label{eq:bcone}
\end{eqnarray}
where we have replaced the Higgs field by its vev $v$.
This looks like a nontrivial boundary condition supplementing
the bulk EOM. Strictly speaking this is not true, since
orbifolds do not have boundaries, only fixed points.
However $Z_2$ orbifold field theories can be thought of as 
simple concatenations of limits of theories with boundaries,
in which case the analogs of (\ref{eq:bcone}) are indeed just
boundary conditions.

This boundary condition tells us that the two charged gauge bosons,
as well as one linear combination of the neutral ones, cannot
have zero modes. From the 4d point of view, the $W^{\pm}$ and $Z$
have eaten the three Goldstone modes of the Higgs to become massive,
as usual. From the 5d point of view, the $W^{\pm}$ and $Z$ are
now massive KK modes! Solving the EOM with the boundary condition
(\ref{eq:bcone}), we get a simple expression for the mass of
these bosons and their heavier KK siblings:
\begin{eqnarray}
M{\rm tan}(ML) = g_5^2v^2 \; .
\label{eq:massf}
\end{eqnarray}
For $v \ll 1/L$, the solution for the lightest KK modes
reduces to the usual expression from the 4d Higgs mechanism:
$M^2 = g_5^2v^2/L = g_4^2v^2$. Remarkably, the solution
of (\ref{eq:massf}) for the lightest massive gauge bosons
also has a smooth limit as $v\rightarrow\infty$:
$M= \pi/2L$.

Thus in this simple 5d gauge theory we have succeeded in taking
a smooth limit where the Higgs boson disappears, but the massive
$W^{\pm}$ and $Z$ bosons remain! This is the simplest example of
a Higgless extra dimensional spontaneously broken gauge 
theory \cite{Csaki:2003dt}.

There is a famous argument \cite{Lee:1977eg}
that in the Higgless Standard Model
the amplitude of elastic scattering of longitudinal massive $W$
bosons blows up at energies around $4\pi M_W/g \sim$
1.8 TeV. From this it is concluded that
one must observe either a Higgs boson or new strong dynamics
at or below this scale. This is the main argument that was used to 
justify building the LHC. Now we see that it is incomplete.

Five dimensional gauge theories of the type I am describing are
happily perturbative up to energy scales of around 
$24\pi^3/g_5^2 = 24\pi^3/g_4^2L$, which is much higher than the
mass of the gauge bosons in the Higgsless limit 
computed from (\ref{eq:massf}). However, these Higgless theories
contain many additional massive KK gauge bosons, which make new
perturbative contributions to $W_LW_L$, $W_LZ$, and $ZZ$ scattering.
These extra KK gauge bosons do precisely the same job usually done
by the Higgs: cancelling the SM contributions which grow with energy like 
$E^4$ and $E^2$. This preserves unitarity and weak coupling up to
a much higher cutoff scale.

Of course these observations are all moot unless one can come up
with realistic Higgless versions of the Standard Model which are
perturbative up beyond 1.8 TeV. Here one
encounters an immediate problem: the enlargement of the electroweak
sector by several KK gauge bosons will generically produce
radiative effects detectable in precision experiments.
Thus generic Higgless models are already excluded by LEP
and our other precision electroweak data.

The best chance for realistic Higgless models seems to come from
Higgless variants of the warped RSI model \cite{Csaki:2003zu}. 
Instead of assuming
that the SM fields are all localized on the TeV brane, we put
the gauge fields and fermions in the bulk. The bulk gauge group 
is taken to be $SU(3)_c\times SU(2)_L \times SU(2)_R \times
U(1)_{B-L}$. A Higgs localized on the Planck brane breaks
$SU(2)_R\times U(1)_{B-L}$ to $U(1)_Y$, while another Higgs
localized on the TeV brane breaks $SU(2)_L\times SU(2)_R$
down to the diagonal $SU(2)$. The combination of these two
breakings is equivalent to the usual SM breaking, preserving
(as in the SM) a custodial isospin which inhibits radiative
corrections to the $\rho$ parameter.

The Higgsless limit of these breakings corresponds to a set
of boundary conditions for the gauge bosons
on the Planck and TeV branes which are
not those of the Randall-Sundrum orbifold. The SM fermions also
obey nontrivial boundary conditions, allowing them to obtain
chiral masses even in the Higgless limit. From the point of
view of $AdS/CFT$ string theory, this means we are introducing
some new kind of $UV$ and $IR$ cutoffs on the $AdS/CFT$ setup,
which may or may not make sense. However at least the bulk gauge
theory is well defined and fully gauge invariant, so as
phenomenologists we are content. 

Unfortunately these warped Higgless models are not realistic
either \cite{Davoudiasl:2003me}-\cite{Csaki:2003sh}.
Perturbative unitarity forces us to make some of the
KK gauge bosons rather light, which in turn generates
contributions to the 
oblique parameters $S$ and $T$ \cite{Peskin:1991sw}
which are
at least twice as large as condoned by the precision data.
This observation has almost been promoted to the status of
a no-go theorem, and has been extended to a large class of
4d Higgless theories using deconstruction. Recently, however,
it was pointed out \cite{Cacciapaglia:2004rb}
that warped Higgless models can be
reconciled with the precision data by tuning parameters of the bulk
fermions in such a way that their couplings to the extra KK
gauge bosons are suppressed. This works, but in turn raises
new issues involving the top and bottom sector, as well as
flavor changing neutral currents. Thus a fully realistic Higgless
model remains elusive.

Nevertheless, the LHC phenomenology of this scenario is interesting
and important. In the case that the couplings of the SM fermions to
the extra KK gauge bosons
{\it are not} very suppressed, this is described 
in \cite{Davoudiasl:2003me}.
The first one or two extra KK copies of the $Z$ should be observable
in Drell-Yan processes at the LHC. The first KK gluon should be
seen as a dijet resonance, and the first KK $W$ should also be
detectable. The difermion KK graviton resonances, which are the
smoking gun of RSI, are not observable in the warped Higgless
models, because the fermions are not localized on the TeV brane
(a possible exception is $gg \rightarrow$ KK graviton
$\rightarrow t_R\bar{t}_R$). KK graviton induced diphoton 
resonances also turn out to be too small to be seen.

Suppose now that viable Higgless models require that the couplings
of the SM fermions to the extra KK gauge bosons {\it are} very
suppressed. This scenario is described in \cite{Birkedal:2004au},
where the authors
ask what can be seen at the LHC in what appears to be
the worst-case scenario. The LHC experiments will see no Higgs,
no supersymmetry (warped models solve the hierarchy problem without
invoking weak scale SUSY), and no new strong interactions (since the theory
is perturbative up to $\sim 10$ TeV). The extra KK gauge bosons are
not visible in Drell-Yan or dijets, because their couplings to
fermions are too weak. Even $W_LW_L$ scattering is not sufficiently
accessible to be conclusive, due to large backgrounds.

What remains in this worst-case analysis is resonant $WZ$
production. This is a completely generic feature of Higgless models,
since $WZ\rightarrow WZ$ must include $s$-channel exchanges of
the extra KK charged gauge bosons, in order to preserve perturbative
unitarity. The $WZ$ intitial state arises from quark bremsstrahlung.
The enhanced golden final state is trileptons + MET + two forward jets.
The estimated KK mass reach is $\sim 1$ TeV with 60 $fb^{-1}$. 

\section{AED}

In our discussion of the ADD scenario, we interpreted the 
parameter $M_*$ in the Gauss law relation (\ref{eqn:gaussrel})
as the fundamental quantum gravity scale of the higher dimensional
theory. We argued that this parameter could be on the order of a
TeV. However even if there are one or more large extra dimensions, it
seems likely (from string theory if nothing else) that there are
additional compactified dimensions on smaller scales. Suppose we
let $R$ denote the size of $n$ large extra dimensions, and $r$ denote
the size of $m$ smaller extra dimensions. Then (\ref{eqn:gaussrel})
should be supplemented by the relation
\begin{eqnarray}
M_*^{n+2} = M^{n+m+2}r^m \; ,
\end{eqnarray}
which relates $M_*$ to the actual quantum gravity scale $M$.
Note that now both $M_*$ and $M_{\rm Planck}$ are derived
parameters, and neither one corresponds to an energy threshold for
new physics.

In the spirit of ADD we take the most optimistic scenario, where
$n=1$ and $m=5$, with $R \simeq 1$ mm and $1/r = $TeV. Then the
quantum gravity scale $M$ is 100-200 TeV. We had better assume that
SM fields are confined to a brane which does not extend in the millimeter
size extra dimension. However there is no reason why some or all of the
SM fields cannot extend in one or more of the TeV$^{-1}$ size extra
dimensions. A 100 TeV quantum gravity scale still leaves the SM with
a little hierarchy problem, but a 100 TeV quantum gravity (and string)
scale is probably more realistic anyway than the TeV assumption of ADD.

This scenario \cite{AED}
is known as asymmetrical extra dimensions (AED).
Even from its acronym we can tell that it is some kind of hybrid of
ADD and UED. If we assume that all of the SM fields propagate in
one TeV$^{-1}$ size extra dimension, and that
this extra dimension is a $Z_2$ orbifold of a circle,
then AED becomes the simple UED model we described above,
with an extra millimeter size hidden dimension added \cite{Macesanu:2002db}.

The original AED model \cite{AED}
however, assumes instead that the SM
fermions are confined to a single 3-brane at $y=0$, 
while the SM gauge bosons
propagate in the bulk of a $Z_2$ orbifolded circle. We are agnostic
about what the Higgs does.
As in UED,
the gauge boson self-couplings will conserve KK parity. However
the couplings of the gauge bosons to the fermions violate conservation
of KK parity. This is because we have placed the fermions asymmetrically
with respect to the translation $y\rightarrow y+r$ which interchanges the
two orbifold fixed points. As a result there is a tree level coupling between
two quarks and the lightest massive KK mode of each gauge boson.

Thus simple AED predicts extra KK copies of the $W$, $Z$, photon,
gluon, and graviton. The effects of KK gravitons are suppressed by
at least $E^4/M^4$, where $E$ is the subprocess energy, so these are
not detectable at the LHC. The extra KK copies of the electroweak
gauge bosons will affect the precision electroweak observables, as
we have already discussed. Current electroweak data already constrains
the AED compactification scale $1/r$ to be greater than about 2 TeV.

The smoking gun of AED is its effect on dijet production 
at the LHC \cite{Dicus:2000hm}.
Tree level single exchanges of virtual KK gluons enhance the production
of dijets with large invariant mass, from quark initial states.
This enhancement can be reliably computed, because for a single
tower of KK gluon states the sum over virtual exchanges is rapidly
convergent. The enhancement is slightly offset by new logarithmically
divergent loop diagrams, which cause the strong coupling $\alpha_s$
to run more rapidly toward asymptotic freedom.

\begin{figure}
\centerline{\epsfxsize 3.0 truein \epsfbox {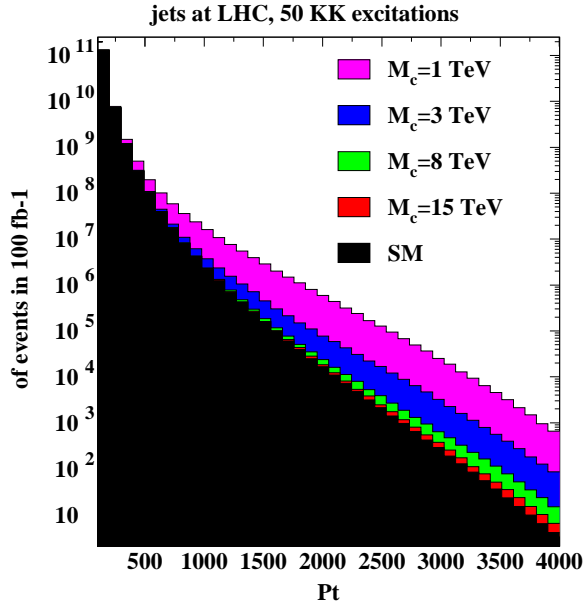}}
\caption{\label{fig:dijet} Number of dijet events versus
the dijet $p_T$, including the enhancement over
the SM rate from virtual KK gluon production. $M_c$ is the same
as the AED scale $1/r$. Taken from reference \cite{Allanach:2004ub}.
\hbox to302pt{}} 
\end{figure}

As can be seen in Figure \ref{fig:dijet}, LHC experiments will be
sensitive to AED for compactification scales as high as
15 TeV. The signal is a smooth excess of high mass dijets.
It is only necessary to analyze a convenient kinematic region
with reasonable statistics, such as 1 TeV $\le m_{jj} \le$ 4 TeV.
Of course this is still a very significant experimental challenge,
including the difficulty of disambiguating this signal from
pdf uncertainties as well as other candidates for new physics.
Note that dijet resonance peaks from on-shell production of KK
gluons are not observable, due to the large width and low rate.

\section{Whatever happened to leptoquarks?}

No survey of physics beyond the Standard Model would be complete
without the mention of leptoquarks. There continue to be
diligent experimental searches for leptoquarks at energy frontier colliders.
However, since the brief excitement at HERA some years ago \cite{hera_lq_excess}, 
the theory community appears to have lost interest. A brief survey
of the literature only turned up three theory papers on leptoquarks from
the past five years. For comparision, this is about 0.1\% of the number
of theory papers written on extra dimensions during the same period.
Why have the theorists forgotten leptoquarks? Should our papers on
leptoquarks be consigned to the same archive as theorist's models for
the high-y anomaly?

The answer to the second question is: not yet. To see why, we need to
be precise about what leptoquarks are. We will define a leptoquark to
be any boson which decays into a lepton and quark through a renormalizable
coupling that respects all SM gauge symmetries.
This means that the squarks of weak scale supersymmetry are leptoquarks,
provided only that we do not completely suppress all of the standard
R parity violating couplings of the form:
\begin{eqnarray}
\lambda^{\prime}_{ijk}\ell_iq_j{\tilde{d}}_k +
\lambda^{\prime}_{ijk}\ell_i{\tilde{q}}_jd_k \quad ,
\end{eqnarray}
where $\ell$, $q$ and $d$ are SM lepton doublets, quark doublets, and
down-type weak singlet quarks, while $\tilde{q}$ and $\tilde{d}$ are
squarks. The labels $i$, $j$, $k$ run over the three generations.

So R parity violating squarks are precisely leptoquarks. Furthermore,
their masses are naturally around the TeV scale, provided that SUSY
has something to do with stabilizing the electroweak scale.

The original motivation for leptoquarks (in the modern era) came from
grand unified theories (GUTs). In GUTs the SM quarks and leptons are
members of the same gauge multiplets, e.g. the $\bar{5}$ and $10$ of 
$SU(5)$, the 16 of $SO(10)$, or the 27 of $E_6$. It follows that some
of the heavy bosons in other GUT representations, such as the $X$, $Y$
gauge bosons of $SU(5)$, or scalars from extended GUT Higgs multiplets,
will be leptoquarks. This is certainly a well-motivated source of
leptoquarks. However the natural mass scale for such leptoquarks is
the unification scale, not the TeV scale.

One way to motivate TeV mass leptoquarks from GUTs is to invoke
extra dimensions. In particular, we construct an $SU(5)$ GUT variation
of the Randall-Sundrum warped geometry \cite{Goldberger:2002pc}.
In this warped GUT the $SU(5)$
gauge bosons and scalars propagate in the bulk, but the fermions are
confined to the Planck brane. Boundary conditions are chosen at the
Planck brane that break $SU(5)$ down to the SM gauge group. These
boundary conditions remove the zero modes of the $SU(5)$ $X$ and $Y$
gauge bosons. There are also TeV mass KK modes of the $X$ and $Y$, but
these are not leptoquarks because the boundary conditions kill
their tree level couplings to SM fermions.

This is an interesting warped GUT in its own right, but let us now
modify it to contain leptoquarks. All we have to do is add some
additional bulk scalars in the $5+\bar{5}$ of $SU(5)$.
We then choose as boundary conditions that their wavefunctions
vanish at the TeV brane. This removes their zero modes, leaving
TeV mass KK modes. These have unsuppressed tree level couplings
to SM fermions. The color triplet weak singlet scalars in these
multiplets are TeV mass leptoquarks. To avoid rapid proton decay,
we can require that they only couple to the third generation.

\section{Beyond Beyond}

The theory community, especially in the previous few years, has
managed to exist in a remarkably spread out superposition of
states in BSM theory space. I am confident that LHC experiments will
collapse this wavefunction, and guide us towards the eigenstate that
Nature has chosen.

\bigskip

{\small Fermilab is operated by Universities Research Association Inc.\ under
Contract No.\ DE-AC02-76CH03000 with the U.S.\ Department of Energy.
The author is grateful for the hospitality of the 
University of Valencia, where this review was completed.
Thanks to M. Spiropulu for helpful comments on the manuscript.}



\bigskip

\begin{thebibliography}{99}

\bibitem{Chung:2003fi}
For a recent review, see
D.~J.~H.~Chung, L.~L.~Everett, G.~L.~Kane, S.~F.~King, J.~Lykken and L.~T.~Wang,
  Phys.\ Rept.\  {\bf 407}, 1 (2005)
  [arXiv:hep-ph/0312378].

\bibitem{Csaki:2003dt}
C.~Csaki, C.~Grojean, H.~Murayama, L.~Pilo and J.~Terning,
Phys.\ Rev.\ D {\bf 69}, 055006 (2004)
[arXiv:hep-ph/0305237].

\bibitem{Harnik:2003rs}
R.~Harnik, G.~D.~Kribs, D.~T.~Larson and H.~Murayama,
Phys.\ Rev.\ D {\bf 70}, 015002 (2004)
[arXiv:hep-ph/0311349].

\bibitem{Kane:2004hm}
G.~L.~Kane, P.~Kumar, J.~D.~Lykken and T.~T.~Wang,
arXiv:hep-ph/0411125.

\bibitem{Marchesano:2004xz}
  F.~Marchesano and G.~Shiu,
  JHEP {\bf 0411}, 041 (2004)
  [arXiv:hep-th/0409132].

\bibitem{Lust:2004dn}
  D.~Lust, S.~Reffert and S.~Stieberger,
  arXiv:hep-th/0410074.

\bibitem{Cvetic:2004xx}
  M.~Cvetic and T.~Liu,
  arXiv:hep-th/0409032.

\bibitem{Camara:2004jj}
  P.~G.~Camara, L.~E.~Ibanez and A.~M.~Uranga,
  Nucl.\ Phys.\ B {\bf 708}, 268 (2005)
  [arXiv:hep-th/0408036].

\bibitem{Kobayashi:2004ya}
  T.~Kobayashi, S.~Raby and R.~J.~Zhang,
  Nucl.\ Phys.\ B {\bf 704}, 3 (2005)
  [arXiv:hep-ph/0409098].

\bibitem{UED}
T.~Appelquist, H.~C.~Cheng and B.~A.~Dobrescu,
Phys.\ Rev.\ {\bf D64} (2001) 035002
[arXiv:hep-ph/0012100].

\bibitem{Cheng:2002ej}
  H.~C.~Cheng, J.~L.~Feng and K.~T.~Matchev,
  Phys.\ Rev.\ Lett.\  {\bf 89}, 211301 (2002)
  [arXiv:hep-ph/0207125].

\bibitem{Servant:2002aq}
  G.~Servant and T.~M.~P.~Tait,
  Nucl.\ Phys.\ B {\bf 650}, 391 (2003)
  [arXiv:hep-ph/0206071].

\bibitem{Cheng:2002ab}
  H.~C.~Cheng, K.~T.~Matchev and M.~Schmaltz,
  Phys.\ Rev.\ D {\bf 66}, 056006 (2002)
  [arXiv:hep-ph/0205314].

\bibitem{Barr:2004ze}
  A.~J.~Barr,
  Phys.\ Lett.\ B {\bf 596}, 205 (2004)
  [arXiv:hep-ph/0405052].

\bibitem{Buras:2003mk}
  A.~J.~Buras, A.~Poschenrieder, M.~Spranger and A.~Weiler,
  Nucl.\ Phys.\ B {\bf 678}, 455 (2004)
  [arXiv:hep-ph/0306158].

\bibitem{Arkani}
N.~Arkani-Hamed, S.~Dimopoulos and G.~Dvali,
Phys.\ Lett.\  {\bf B429} (1998) 263;
Phys.\ Rev.\  {\bf D59} (1999) 086004.

\bibitem{Lykken:1998ec}
  J.~Lykken, E.~Poppitz and S.~P.~Trivedi,
  Nucl.\ Phys.\ B {\bf 543}, 105 (1999)
  [arXiv:hep-th/9806080].

\bibitem{Antoniadis:1990ew}
I.~Antoniadis,
Phys.\ Lett.\ B {\bf 246}, 377 (1990).

\bibitem{Lykken:1996fj}
J.~D.~Lykken,
Phys.\ Rev.\ D {\bf 54}, 3693 (1996)
[arXiv:hep-th/9603133].

\bibitem{Giudice:1998ck}
G.~F.~Giudice, R.~Rattazzi and J.~D.~Wells,
Nucl.\ Phys.\ {\bf B544} (1999) 3 [arXiv:hep-ph/9811291].

\bibitem{Han:1998sg}
T.~Han, J.~D.~Lykken and R.~J.~Zhang,
Phys.\ Rev.\ D {\bf 59} (1999) 105006 [arXiv:hep-ph/9811350].

\bibitem{Randall:1999ee}
L.~Randall and R.~Sundrum,
Phys.\ Rev.\ Lett.\  {\bf 83}, 3370 (1999)
[arXiv:hep-ph/9905221].

\bibitem{RSII}
L.~Randall and R.~Sundrum,
Phys.\ Rev.\ Lett.\  {\bf 83}, 4690 (1999)
[arXiv:hep-th/9906064].

\bibitem{Davoudiasl:1999jd}
H.~Davoudiasl, J.~L.~Hewett and T.~G.~Rizzo,
Phys.\ Rev.\ Lett.\ {\bf 84}, 2080 (2000)
[arXiv:hep-ph/9909255].

\bibitem{Hewett-Smaria}
J.~Hewett and M.~Spiropulu,
Ann.\ Rev.\ Nucl.\ Part.\ Sci.\ {\bf 52}, 397 (2002)
[arXiv:hep-ph/0205106].

\bibitem{Arkani-Hamed:2002qy}
N.~Arkani-Hamed, A.~G.~Cohen, E.~Katz and A.~E.~Nelson,
JHEP {\bf 0207}, 034 (2002)
[arXiv:hep-ph/0206021].

\bibitem{Arkani-Hamed:2002qx}
N.~Arkani-Hamed, A.~G.~Cohen, E.~Katz, A.~E.~Nelson, T.~Gregoire and J.~G.~Wacker,
JHEP {\bf 0208}, 021 (2002)
[arXiv:hep-ph/0206020].

\bibitem{Gregoire:2002ra}
T.~Gregoire and J.~G.~Wacker,
JHEP {\bf 0208}, 019 (2002)
[arXiv:hep-ph/0206023].

\bibitem{Low:2002ws}
I.~Low, W.~Skiba and D.~Smith,
Phys.\ Rev.\ D {\bf 66}, 072001 (2002)
[arXiv:hep-ph/0207243].

\bibitem{amsterdam}
M.~Schmaltz,
Nucl.\ Phys.\ Proc.\ Suppl.\  {\bf 117}, 40 (2003)
[arXiv:hep-ph/0210415].

\bibitem{kaplanschmaltz}
D.~E.~Kaplan and M.~Schmaltz,
JHEP {\bf 0310}, 039 (2003)
[arXiv:hep-ph/0302049].

\bibitem{Chang:2003un}
S.~Chang and J.~G.~Wacker,
Phys.\ Rev.\ D {\bf 69}, 035002 (2004)
[arXiv:hep-ph/0303001].

\bibitem{skibaterning}
W.~Skiba and J.~Terning,
Phys.\ Rev.\ D {\bf 68}, 075001 (2003)
[arXiv:hep-ph/0305302].

\bibitem{Chang:2003zn}
S.~Chang,
JHEP {\bf 0312}, 057 (2003)
[arXiv:hep-ph/0306034].

\bibitem{Kaplan:2004cr}
D.~E.~Kaplan, M.~Schmaltz and W.~Skiba,
Phys.\ Rev.\ D {\bf 70}, 075009 (2004)
[arXiv:hep-ph/0405257].

\bibitem{simplest}
M.~Schmaltz,
JHEP {\bf 0408}, 056 (2004)
[arXiv:hep-ph/0407143].


\bibitem{Cheng:2004yc}
H.~C.~Cheng and I.~Low,
JHEP {\bf 0408}, 061 (2004)
[arXiv:hep-ph/0405243].

\bibitem{Low:2004xc}
I.~Low,
JHEP {\bf 0410}, 067 (2004)
[arXiv:hep-ph/0409025].


\bibitem{Hubisz:2004ft}
  J.~Hubisz and P.~Meade,
  Phys.\ Rev.\ D {\bf 71}, 035016 (2005)
  [arXiv:hep-ph/0411264].



\bibitem{Csaki:2002qg}
C.~Csaki, J.~Hubisz, G.~D.~Kribs, P.~Meade and J.~Terning,
Phys.\ Rev.\ D {\bf 67}, 115002 (2003)
[arXiv:hep-ph/0211124].

\bibitem{Hewett:2002px}
J.~L.~Hewett, F.~J.~Petriello and T.~G.~Rizzo,
JHEP {\bf 0310}, 062 (2003)
[arXiv:hep-ph/0211218].

\bibitem{Csaki:2003si}
C.~Csaki, J.~Hubisz, G.~D.~Kribs, P.~Meade and J.~Terning,
Phys.\ Rev.\ D {\bf 68}, 035009 (2003)
[arXiv:hep-ph/0303236].

\bibitem{Chivukula:2002ww}
R.~S.~Chivukula, N.~J.~Evans and E.~H.~Simmons,
Phys.\ Rev.\ D {\bf 66}, 035008 (2002)
[arXiv:hep-ph/0204193].

\bibitem{Chen:2003fm}
M.~C.~Chen and S.~Dawson,
Phys.\ Rev.\ D {\bf 70}, 015003 (2004)
[arXiv:hep-ph/0311032].

\bibitem{Casalbuoni:2003ft}
R.~Casalbuoni, A.~Deandrea and M.~Oertel,
JHEP {\bf 0402}, 032 (2004)
[arXiv:hep-ph/0311038].


\bibitem{Azuelos:2004dm}
G.~Azuelos {\it et al.},
arXiv:hep-ph/0402037.

\bibitem{Han:2003wu}
T.~Han, H.~E.~Logan, B.~McElrath and L.~T.~Wang,
Phys.\ Rev.\ D {\bf 67}, 095004 (2003)
[arXiv:hep-ph/0301040].

\bibitem{Burdman:2002ns}
G.~Burdman, M.~Perelstein and A.~Pierce,
Phys.\ Rev.\ Lett.\  {\bf 90}, 241802 (2003)
[Erratum-ibid.\  {\bf 92}, 049903 (2004)]
[arXiv:hep-ph/0212228].

\bibitem{Allanach:2004ub}
B.~C.~Allanach {\it et al.}  [Beyond the Standard Model Working Group
Collaboration],
arXiv:hep-ph/0402295.


\bibitem{Lee:1977eg}
B.~W.~Lee, C.~Quigg and H.~B.~Thacker,
Phys.\ Rev.\ D {\bf 16}, 1519 (1977).

\bibitem{Csaki:2003zu}
C.~Csaki, C.~Grojean, L.~Pilo and J.~Terning,
Phys.\ Rev.\ Lett.\  {\bf 92}, 101802 (2004)
[arXiv:hep-ph/0308038].

\bibitem{Davoudiasl:2003me}
H.~Davoudiasl, J.~L.~Hewett, B.~Lillie and T.~G.~Rizzo,
Phys.\ Rev.\ D {\bf 70}, 015006 (2004)
[arXiv:hep-ph/0312193].

\bibitem{Hewett:2004dv}
J.~L.~Hewett, B.~Lillie and T.~G.~Rizzo,
JHEP {\bf 0410}, 014 (2004)
[arXiv:hep-ph/0407059].

\bibitem{Barbieri:2003pr}
R.~Barbieri, A.~Pomarol and R.~Rattazzi,
Phys.\ Lett.\ B {\bf 591}, 141 (2004)
[arXiv:hep-ph/0310285].

\bibitem{Chivukula:2004pk}
R.~S.~Chivukula, E.~H.~Simmons, H.~J.~He, M.~Kurachi and M.~Tanabashi,
Phys.\ Rev.\ D {\bf 70}, 075008 (2004)
[arXiv:hep-ph/0406077].

\bibitem{Chivukula:2004af}
R.~S.~Chivukula, E.~H.~Simmons, H.~J.~He, M.~Kurachi and M.~Tanabashi,
Phys.\ Lett.\ B {\bf 603}, 210 (2004)
[arXiv:hep-ph/0408262].

\bibitem{Georgi:2004iy}
H.~Georgi,
arXiv:hep-ph/0408067.

\bibitem{Csaki:2003sh}
C.~Csaki, C.~Grojean, J.~Hubisz, Y.~Shirman and J.~Terning,
Phys.\ Rev.\ D {\bf 70}, 015012 (2004)
[arXiv:hep-ph/0310355].

\bibitem{Peskin:1991sw}
M.~E.~Peskin and T.~Takeuchi,
Phys.\ Rev.\ D {\bf 46}, 381 (1992).

\bibitem{Cacciapaglia:2004rb}
G.~Cacciapaglia, C.~Csaki, C.~Grojean and J.~Terning,
arXiv:hep-ph/0409126.

\bibitem{Birkedal:2004au}
  A.~Birkedal, K.~Matchev and M.~Perelstein,
  arXiv:hep-ph/0412278.


\bibitem{AED}
J.~Lykken and S.~Nandi, Phys.\ Lett.\ {\bf B485} (2000) 224
[arXiv:hep-ph/9908505].

\bibitem{Macesanu:2002db}
C.~Macesanu, C.~D.~McMullen and S.~Nandi,
Phys.\ Rev.\ D {\bf 66}, 015009 (2002)
[arXiv:hep-ph/0201300].

\bibitem{Dicus:2000hm}
D.~A.~Dicus, C.~D.~McMullen and S.~Nandi,
Phys.\ Rev.\ D {\bf 65}, 076007 (2002)
[arXiv:hep-ph/0012259].


\bibitem{hera_lq_excess}
C.~Adloff {\it et al.}  [H1 Collaboration],
Eur.\ Phys.\ J.\ C {\bf 11}, 447 (1999)
[Erratum-ibid.\ C {\bf 14}, 553 (1999)]
[hep-ex/9907002]; \\
J.~Breitweg {\it et al.}  [ZEUS Collaboration],
Eur.\ Phys.\ J.\ C {\bf 16}, 253 (2000)
[hep-ex/0002038].

\bibitem{Goldberger:2002pc}
  W.~D.~Goldberger, Y.~Nomura and D.~R.~Smith,
  Phys.\ Rev.\ D {\bf 67}, 075021 (2003)
  [arXiv:hep-ph/0209158].

\end{thebibliography}
\end{document}